# Near-real-time diagnosis of electron optical phase aberrations in scanning transmission electron microscopy using an artificial neural network

Giovanni Bertoni,[1*] Enzo Rotunno,[1*] Daan Marsmans,[2] Peter Tiemeijer,[2] Amir H. Tavabi,[3] Rafal E. Dunin-Borkowski,[3] and Vincenzo Grillo[1]

[1.] Istituto Nanoscienze, Consiglio Nazionale delle Ricerche, Via G. Campi 213/A, 41125 Modena, Italy.

[2.] Thermo Fisher Scientific, PO Box 80066, 5600 KA Eindhoven, the Netherlands.

[3.] Ernst Ruska-Centre for Microscopy and Spectroscopy with Electrons, Forschungszentrum Jülich, 52425 Jülich, Germany.

*Corresponding authors: giovanni.bertoni@cnr.it, enzo.rotunno@cnr.it

**Abstract**

The key to optimizing spatial resolution in a state-of-the-art scanning transmission electron microscope is the ability to precisely measure and correct for electron optical aberrations of the probe-forming lenses. Several diagnostic methods for aberration measurement and correction with maximum precision and accuracy have been proposed, albeit often at the cost of relatively long acquisition times. Here, we illustrate how artificial intelligence can be used to provide near-real-time diagnosis of aberrations from individual Ronchigrams. The demonstrated speed of aberration measurement is important as microscope conditions can change rapidly, as well as for the operation of MEMS-based hardware correction elements that have less intrinsic stability than conventional electromagnetic lenses.

**1 Introduction**

Aberration correction has been key to achieving incredible spatial resolution using state-of-the-art transmission electron microscopes (TEMs) [1-4]. At the same time, it has required great effort in hardware fabrication and instrument control. In order to circumvent the unavoidable spherical aberration of round electromagnetic lenses and residual non-symmetrical aberrations, many lenses and multipoles have to be combined and controlled with extreme precision [5-9]. The situation is similar to that of adaptive correction of aberrations and atmospheric effects in astronomical telescopes, for which quick adjustments are required to account for atmospheric changes [10, 11].

The conceptual scheme that underlies control of aberration correction is based on decomposition of the complex phase landscape that acts on an electron in a lens into well-known Zernike polynomials [12], which can be expressed in the form of a classical aberration polynomial series. Aberration correction aims to introduce a similar series of polynomials that have opposite signs, in order to provide near-perfect compensation up to a well-defined order. An experimental microscopist is typically trained to recognise and to manually correct lower order aberrations, such as defocus ($C_1$) and astigmatism ($A_1$). However, in order to obtain the best possible resolution of the microscope, higher-order aberrations must also be tuned. This procedure is too difficult to complete manually and requires sophisticated automated procedures for diagnosing and correcting the aberrations. Aberration diagnosis plays a crucial role and its practical implementation is a key factor affecting the precision and speed of correction. Many ideas and concepts have been introduced to make aberration evaluation as reliable as possible for both the imaging system after the sample and the probe forming system before it.

For the imaging system, the gold standard is the use of a Zemlin tableau [13], where different inclined plane wave illuminations are used to map variations in lower order aberrations such as apparent defocus and astigmatism. In the case of scanning TEM (STEM), the concept is similar. For example, a convenient approach involves measurements of beam shift [14] (the lowest order effect) or changes in contrast in STEM imaging as a function of tilt [15].

For the work presented in this paper, the most informative approach involves the analysis of a Ronchigram, i.e., a coherent diffraction image of a convergent probe after scattering through a thin crystalline or amorphous sample [16]. A Ronchigram contains both diffraction information and a distorted image of the sample. The image has a locally-varying magnification, which depends on the second derivative of the aberration phase [17]. Some methods combine information from Ronchigrams recorded at different tilts. However, the Ronchigram itself can be regarded as a superimposed diffraction map for different tilt components of the probe. In this sense, it is similar to a full Zemlin tableau. This correspondence has been used to propose an aberration measurement approach [18]. Lupini et al. demonstrated that, after segmenting a Ronchigram of an amorphous material into many subimages (or tiles), each of which corresponds to a different effective tilt [19], it is possible to apply a defocus/astigmatism fitting routine just as in a Zemlin-like scheme. Alternative aberration diagnosis methods use other interesting of Ronchigrams, which are indicative of the richness of the information that it contains [20-23].

If a single Ronchigram can be used to evaluate aberrations, then it can be used to provide fast correction feedback. This capability can also be useful for MEMS-based optics [24, 25], which can even be used for spherical aberration correction [26], albeit with lower stability than conventional optics.

A recent addition to this complicated landscape of methods and concepts is the introduction of artificial intelligence (AI) and deep learning. AI makes use of a training procedure to recognise patterns and hidden recurrences in a dataset, allowing the parametric behavior of a system to be predicted [25, 27]. A recent example of the use of AI in electron optics, which also involves the use of MEMS technology, is our application of AI to tune an orbital angular momentum (OAM) sorter [28-31]. An OAM sorter is a combination of two tunable phase elements, which disperse an electron beam into a spectrum of discrete OAM states. We applied AI to a single image of an OAM spectrum to determine all relevant alignment parameters and to automatically tune the OAM sorter [24, 25]. Recent approaches have demonstrated the use of deep learning techniques to optimize a STEM aperture for improving resolutio[32, 33], confirming that the analysis of a single Ronchigram [34] can be translated directly into aberration diagnostics.

Here, we provide the first demonstration of aberration diagnosis based on an artificial neural network (ANN). We start from the idea of segmentation discussed by Lupini et al. [19] However, instead of relying on a semi-analytical approach, we apply ANN recognition to the full set of images without assuming any approximation of the sub-image diffractograms.

**2 Methods**

**2.1 Ronchigram modeling**

As a starting point, we describe diffraction information in a Ronchigram in terms of the scattering of a convergent probe through a phase object. The probe is described in real space $x$ as $\psi_p(x) = IFT\{e^{-i\chi(q)}A(q)\}$, where $q$ is a reciprocal space coordinate (*i.e.*, frequency) conjugated to $x$ and $\chi(q)$ is the aberration function of the illumination system and $A(q)$ is a top hat function introduced by the condenser aperture. Hereafter, *IFT* and *FT* refer to inverse and direct Fourier transforms, respectively. The sample has a multiplicative effect, which is described by a transmission function $T(x) = e^{-i\sigma V(x)}$, where the interaction parameter $\sigma = 2\pi m_0 \gamma e \lambda / h^2$, $m_0$ and $e$ are the rest mass and charge of the electron, $\lambda$ is the electron wavelength, $\gamma$ the relativistic factor, $h$ is the Planck constant and $\sigma V(x)$ is the phase shift produced by the sample. If the sample is sufficiently thin, then $V(x)$ is the effective projected atomic potential. The intensity $I(q)$ of the Ronchigram function $\psi_t(q)$ in reciprocal space can be written in the form

$$I(\boldsymbol{q}) = |\psi_t(\boldsymbol{q})|^2 = \left|FT\big(IFT\{e^{-i\chi(\boldsymbol{q})}\}e^{-i\sigma V(\boldsymbol{x})}\big)\right|^2. \tag{1}$$

**2.2 Sample function**

The sample potential $V(x)$ should be as close as possible to the real function that describes the potential of an amorphous phase object (in our case, a thin film of amorphous C, or a-C). The way in which the sample transmits spatial frequencies is critical for ANN training. As a first approach, a random distribution $V(x) = (\bar{V}t)\mathrm{rand}(x)$ can be used, where $\mathrm{rand}(x)$ is a uniform random number in the range [0, 1]. A good choice for $\sigma\bar{V}t$ is the value $\pi/4$, as used by Schnitzer et al. [35], as it is close to the value obtained from the weak phase object (WPO) approximation by considering an interaction constant $\sigma = 0.0065$ V$^{-1}$nm$^{-1}$ at 300 kV, a sample thickness $t \sim 10$ nm and a mean inner potential $\bar{V} = 10.7$ V for a-C [36, 37]. However, in this first approach, the potential contains all frequencies with the same weight (*i.e.*, white noise), which is unrealistic. A better way in which to build the potential is to also consider the dependence on the frequency of transmission of a thin amorphous C film (or the scattering intensity $I_e(q)$) in reciprocal space [38]. A good approximation (assuming azimuthal symmetry of scattering) is given by the electron atomic scattering factor $f(q)$, which can be calculated analytically as a sum of Gaussian functions, as described by Peng et al. [39] Its angular dependence is shown in Fig. 1, in which the electron scattering factor is compared with the square root of the scattered intensity measured from a thin a-C film at 300 kV. Thermal vibrations can be considered as a further multiplicative factor with a Gaussian function $b(q)$ in reciprocal space by using a Debye-Waller factor (here taken to be 0.5 Å$^2$). Both terms act as low pass filters in the frequency domain for $V(x)$.

**2.3 Aberration function**

In the extended notation of Lupini et al. [19], which is derived from the well-established Krivanek notation [40], the aberration function $\chi(q)$ in reciprocal space can be written in the form

$$\chi(\boldsymbol{q}) = 2\pi \sum_n \lambda^n \sum_{m=0}^{n+1} \left[\frac{C_{n,m,a}q^{n+1}\cos(m\phi) + C_{n,m,b}q^{n+1}\sin(m\phi)}{n+1}\right], \tag{2}$$

where $C_{n,m,a}$ and $C_{n,m,b}$ are aberration coefficients of order *n* and symmetry *m*, under the condition that $n + m$ is an odd number. The coefficients reduce to one when $m = 0$ (*e.g.*, for defocus and spherical aberration). In Eq. 2, $q$ is the length of the vector $\boldsymbol{q}$ and $\phi$ is the azimuthal angle (with the optical axis in the center). Most software truncates $\chi(\boldsymbol{q})$ at $n = 5$. We use a similar truncation below, as explained in Section 3.2, as well as the traditional nomenclature for the lower orders (see Table 1).

**2.4 Detector**

The Ronchigram intensity $I(\boldsymbol{q})$ is typically recorded using a digital detector, resulting in convolution a of $I(\boldsymbol{q})$ with the point spread function (PSF) $p$ of the camera:

$$I(\tilde{q}) = \left|IFT\{FT(I(\tilde{q}))FT(p(\tilde{q}))\}\right| + c, \tag{3}$$

where $\tilde{q}$ indicates that the convolution is performed in the pixel dimension. The PSF introduces further attenuation of high spatial frequencies. The constant $c$ takes into account the counting noise of the camera. The absolute value is a reminder that the result of the convolution is real.

An example of the fidelity of the model is presented in Fig. 2, in the form of a comparison of radial profiles of Fourier transforms of an experimental defocused Ronchigram (blue) and a simulated Ronchigram model from Eqs 1-3 (orange) at 300 kV. The intensities of the maxima and minima are well reproduced, as is the decrease in contrast at high spatial frequencies.

**3. Theory and calculations**

The method involves dividing an experimental Ronchigram recorded from an amorphous C film (*e.g.*, a standard C film on a Cu mesh grid) into a set of sub-images and numerically performing a Fourier transform of each sub-image. The results are stacked to form a 3D image, which is fed into the ANN for pattern recognition. Unlike the analytical method of Lupini et al. [19], the fitting procedure of the ANN analyzes the overall pattern and not the separate images.

**3.1 ANN model.** In order to fit Ronchigrams, a custom-made VGG16-type convolutional neural network was implemented in Python using the TensorFlow library. A sketch of the network is shown in Fig. 3a. The ANN is made of 4 sequential blocks, comprising a 2D convolution layer followed by a pooling layer. In order to avoid overfitting, a 20% dropout layer was inserted after every pooling layer. After these blocks, a flatten layer was used, followed by 3 dense layers to output the 8 fitted aberration values needed for calculating $C_{1,0}$, $C_{1,2}$, $C_{2,1}$, $C_{2,3}$, and $C_{3,0}$.

**3.2 ANN training.** A dataset was built with $N$ = 20000 sets of 25 (5 × 5) diffractograms from Ronchigram images (Fig. 3b). Each Ronchigram was calculated at angles up to 40 mrad according to Eq. 1 (with $C_{1,0}$ ~ -2000 nm) and at 2048 × 2048 pixels to reduce the artifacts in the Fourier transforms due to the rapidly varying transmission function $e^{-i\chi(q)}$. Each generated image $I(q)$ was then binned by a factor of 4 and divided into a set of 5 × 5 patterns (or tiles), each of which was 64 × 64 pixels in size (Fig. 2c). The 25 diffractograms from the tiles were then calculated (Fig. 2d). Each set of 25 diffractograms constitutes one of 24000 inputs for ANN training, *i.e.*, the input matrix has dimensions (24000, 64, 64, 25). The corresponding (24000, *j*) matrix has *j* known aberration coefficients. For clarity, in the example shown in Fig. 3b-d the value of $C_{3,0}$ (spherical aberration) is exaggerated to show its effect on the diffractograms ($C_{3,0}$ = 1 mm). ANN training was limited to the *j* = 8 principal aberration values, which were randomly generated in the intervals reported in Table 1. Residual aberrations up to *n* = 5 were generated according to a normal distribution centered around 0 (corrected values) with a standard deviation of 0.1 of their maximum absolute values, which were considered to be 10 μm for *n* = 3 (except for $C_{3,0}$), 100 μm for *n* = 4, and 10 mm for *n* = 5. The network was trained for 200 epochs with a batch size of 40 and a validation set of 20% (4800 sets of diffractograms). After convergence, the loss function was below 0.01 and was evaluated as the mean square error of the 8 fitted parameters (aberration coefficients).

**4 Results and discussion**

**4.1 ANN validation test**

Table 1 shows the mean absolute error (MAE) for the 8 fitted aberration parameters after ANN training. The errors are ~5% of the maximum values of the aberrations in their respective ranges. The ANN is therefore capable of reducing the aberrations by a factor of ~20. This result is promising, considering that only a single image acquisition is needed for the ANN to measure the aberration coefficients. After convergence, the ANN was tested on a freshly-generated new set of Ronchigrams. An example of Ronchigram images for defocus $C_{1,0}$ = 0 with corresponding probe images before and after aberration correction using values measured by the ANN are shown in Fig. 4. A figure of merit is given by the quarter-wave criterion [41, 42], which is based on the region of frequencies $q$ in which the aberration phase $|\chi(q)| \leq \pi/4$. The optimal aperture can then be defined by a circle with radius $\min|\chi(q)| = \pi/4$ corresponding to semi-angle $\alpha$. Based on the aperture $\alpha$, the resolution of the STEM probe can be derived, according to the Rayleigh criterion, to be $d = 0.61(\lambda/\alpha)$. The example shown in Fig. 3 confirms that the optimal aperture $\alpha$ and resolution $d$ are improved dramatically after ANN correction, resulting in a probe that provides atomic spatial resolution.

The precision achieved in the estimations of the aberrations using the ANN is limited (see Table 1). The dominant value limiting the resolution is two-fold astigmatism $C_{1,2}$, while the values for $C_{2,1}$, $C_{2,3}$, and $C_{3,0}$ are in the range expected for sub-Å resolution [43, 44]. In order to reach optimal resolution (*e.g.*, $d < 80$ pm or $\alpha > 15$ mrad at 300 kV), the residual value of $C_{1,2}$ should be kept below 1 nm. Regular re-tuning may be required during an experiment, or it may be possible to use iterative procedures based on contrast enhancement from repeated STEM image acquisitions [45]. The observed lack of precision may result partly from the fact that the amount of information in a single Ronchigram is limited. This limitation may be caused by noise (such as shot noise and camera noise), but also by the fact that some features in a Ronchigram are related to the structure of the sample rather than to the aberrations. A comparison with existing methods based on measurements from a single Ronchigram indicates a similar level of uncertainty [18, 19].

With respect to existing methods, ANN-based diagnosis is very fast, corresponding to ~150 ms from a single image on a standard PC. In addition, in principle it needs only a single image. This time should to be added to that required for acquisition of the image (usually 1–2 s). A simple evolution of the method may therefore involve the use of an iterative approach to minimize the residual values. We tested an iterative procedure for ANN correction on synthetic data by measuring the optimal STEM aperture $\alpha$ according to the quarter-wave criterion at each iteration. During this procedure, the values of the aberrations were not fully corrected. Instead, only a fraction $R$ of the calculated correction was applied each time. We chose a correction factor $R = 1 - \alpha/\alpha_0$, where $\alpha_0$ is the target optimal convergence, set for instance to $\alpha_0 = 12.1$ mrad, which corresponds to $d_0 < 100$ pm at 300 kV. By using this criterion, the ANN typically reached convergence of correction in 1-5 iterations (*i.e.*, 5-10 s including acquisition time). This is a fast procedure for reaching an acceptable resolution value online during imaging. For better tuning towards 60-80 pm, more precise (manual or assisted) correction of $A_1$ is needed. It is unclear if AI routines can achieve this level of precision from a single Ronchigram. In addition, during experiments, $C_{1,0}$ needs continuous adjustment due to differences in the height and thickness of the sample at different positions, while $C_{1,2}$ can also be tuned manually or using standard routines.

Future improvements could be achieved, for instance, by increasing the training dataset substantially (by at least a factor 10) or by increasing the complexity and topology of the ANN.

Another possibility would be to test different segmentation strategies of the Ronchigrams into sub-images, such as increasing the number of patterns (*e.g.*, from 5 × 5 to 7 × 7) or mixing different segmentation patterns (*e.g.*, 5 × 5 and 4 × 4). However, these techniques require an increase in hardware memory requirements and computation time for generating the training dataset and for achieveing ANN convergence.

**4.2 ANN results on experimental data**

Several series of Ronchigrams were recorded at 300 kV on a Thermo Fisher Scientific Spectra TEM equipped with a probe aberration corrector from CEOS GmbH. A convergence aperture of 64 mrad was used and 2048 × 2048 images were acquired with a Ceta camera (Thermo Fisher Scientific). The aberrations $C_{1,0}$, $C_{1,2}$, $C_{2,1}$, $C_{2,3}$, and $C_{3,0}$ were mistuned intentionally in a broad range of combinations and measured for comparison using the built-in software from CEOS GmbH. The mistuned aberrations were also quantified from a single Ronchigram using the ANN. As a benchmark, the ANN results we recompared with values of the aberrations measured using the standard Zemlin tableau method of the probe corrector. The Zemlin tableau requires approximately 10 sets of STEM images recorded at two defocus values by tilting the probe, resulting in very precise aberration measurements at the expense of a long acquisition time (30-60 s). Figure 5 shows the results obtained from the ANN for the aberrations $C_{1,2}$, $C_{2,1}$, $C_{2,3}$ and $C_{3,0}$. The values measured using the ANN are close to the Zemlin tableau values considering 95% confidence intervals. Even the simple fact that the ANN determines the signs of the aberrations correctly is already a good result. It means indeed that the ANN can be used to adjust the aberrations iteratively by assuring convergence. We expect that it can be improved further by continuing the training online on experimental Ronchigram images directly at the microscope. A possible future implementation may involve training the ANN on crystalline materials [22, 23] (either experimental or calculated) so that the AI discerns aberrations directly for a sample under investigation without the need to move the probe onto an amorphous film, thereby reducing the acquisition time and the risk of drift of aberration values.

**5. Conclusions**

We have demonstrated that an ANN can be used to measure aberrations from Ronchigrams for fast online tuning, offering an approach that can be used to keep a STEM tuned and assist an operator. Correction to ~5% of the uncorrected values was demonstrated on synthetic data, corresponding to a spatial resolution of better than 100 pm from a few sets of 1-5 Ronchigrams of an amorphous C film. Despite the poor precision of the ANN when the aberrations are close to zero, the speed of the ANN makes it valuable for tuning and continuous diagnosis of dominant aberrations during high-resolution STEM directly at the microscope.

## 6. Acknowledgments

The authors acknowledge funding from the European Union's Horizon 2020 Research and Innovation Programme under Grant Agreement No. 964591 'SMART-electron'. This project has received funding from the European Union's Horizon 2020 Research and Innovation Programme (Grant No. 823717, project 'ESTEEM3').

## Appendix A. Relations between coefficients

A simple relation between the present notation using two coefficients $(C_{n,m,a}, C_{n,m,b})$ for non-symmetrical aberrations and the standard notation using one coefficient and the azimuthal angle $(C_{n,m}, \phi_{n,m})$ can be derived starting from Krivanek notation:

$$\chi(\mathbf{q}) = 2\pi \sum_n \lambda^n \sum_{m=0}^{n+1} \left[ \frac{C_{n,m} \theta^{n+1} \cos(m(\phi - \phi_{n,m}))}{n+1} \right]. \qquad (A.1)$$

The trigonometric relation

$$C_{n,m} \cos\left(m(\phi - \phi_{n,m})\right) = C_{n,m} \cos(m\phi)\cos(m\phi_{n,m}) + C_{n,m}\sin(m\phi)\sin(m\phi_{n,m}), \qquad (A.2)$$

provides the following relationship between the coefficients and the angle $\phi_{n,m}$ :

$$C_{n,m} \cos(m\phi_{n,m}) = C_{n,m,a}, \qquad (A.3.1)$$

$$C_{n,m} \sin(m\phi_{n,m}) = C_{n,m,b}. \qquad (A.3.2)$$

Equations A.3.1 and A.3.2 can be rewritten in the form:

$$C_{n,m} = \sqrt{C_{n,m,a}^2 + C_{n,m,b}^2}, \qquad (A.4.1)$$

$$\phi_{n,m} = \frac{1}{m} \tan^{-1}\left(\frac{C_{n,m,b}}{C_{n,m,a}}\right). \qquad (A.4.2)$$

**Tables and Figures**

| Aberration | Symbol | Coefficients (this work) | | Range | MAE | | % Error | |
|---|---|---|---|---|---|---|---|---|
| Defocus | $C_1$ | $C_{1,0}$ | | [-2200 nm, -1800 nm] | 15 nm | | 0.8 | |
| Spherical aberration | $C_3$ | $C_{3,0}$ | | [-100 μm, 100 μm] | 3.49 μm | | 3.5 | |
| 2-fold astigmatism | $A_1$ | $C_{1,2,a}$ | $C_{1,2,b}$ | [-100 nm, 100 nm] | 4.7 nm | 4.8 nm | 4.7 | 4.8 |
| Coma | $B_2$ | $C_{2,1,a}$ | $C_{2,1,b}$ | [-1000 nm, 1000 nm] | 49 nm | 48 nm | 4.9 | 4.8 |
| 3-fold astigmatism | $A_2$ | $C_{2,3,a}$ | $C_{2,3,b}$ | [-1000 nm, 1000 nm] | 37 nm | 39 nm | 3.7 | 3.9 |

**Table 1.** Mean absolute error (MAE) in each of the 8 principal aberrations after artificial neural network training. The aberrations were generated randomly in the intervals reported in the 'Range' column. The relative error (%) was obtained by dividing the MAE by the maximum value of the corresponding aberration. The observed higher precision in $C_{1,0}$ is expected, as the defocus is related simply to the position of the first minimum in the diffractogram. The other aberrations depend on the shapes of the rings and are characterized by similar relative errors <5%.

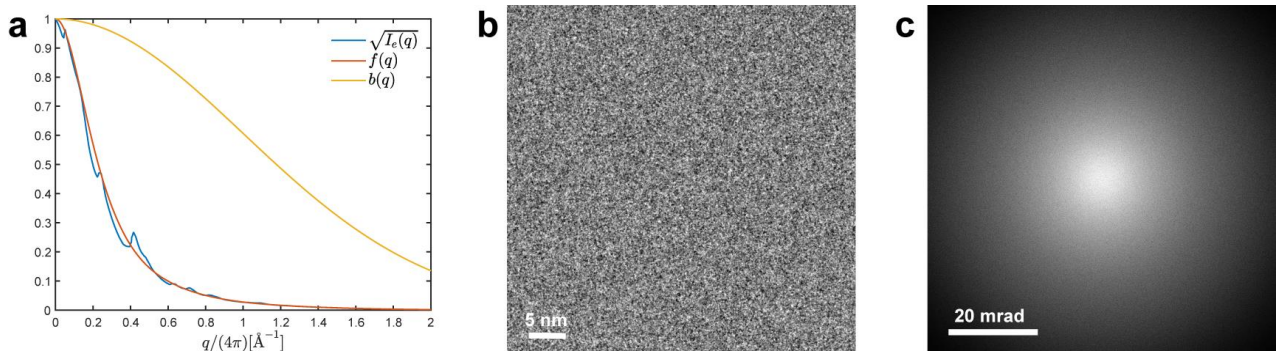

**Figure 1.** (a) Plot of the square root of the intensity of elastic scattering from a thin film of a-C, as measured from electron diffraction (blue curve) [38] and compared with an analytical function for the atomic scattering factor of carbon $f(q)$ (red curve) [39] and the scattering profile expected from thermal vibrations $b(q)$ (orange curve). (b) Calculated $V(\boldsymbol{x})$ after taking into account the scattering described in (a). (c) Fourier transform $FT(V(\boldsymbol{x}))$.

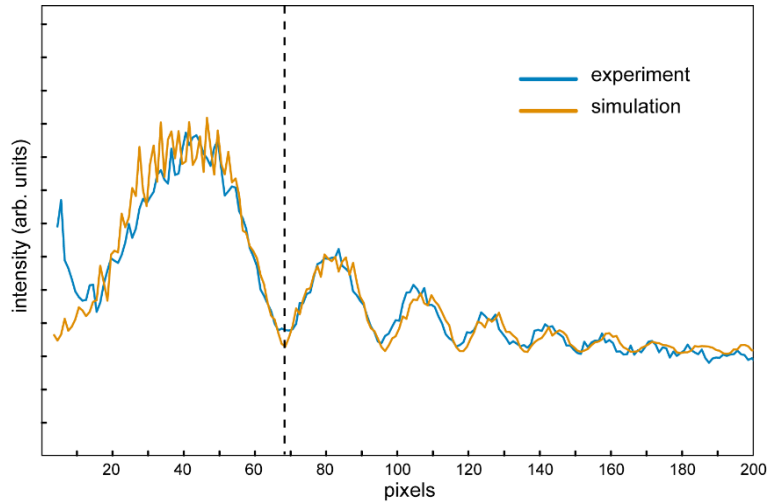

**Figure 2.** Comparison of radial profiles of the Fourier transforms of an experimental defocused Ronchigram (blue curve) and a simulated model according to Eqs. 1-3 (orange curve) at 300 kV. The other aberration values are set to zero. The defocus $C_{1,0}$ can be calculated from the first minimum in the profile ($C_{1,0}$ = −1890 nm, dashed line).

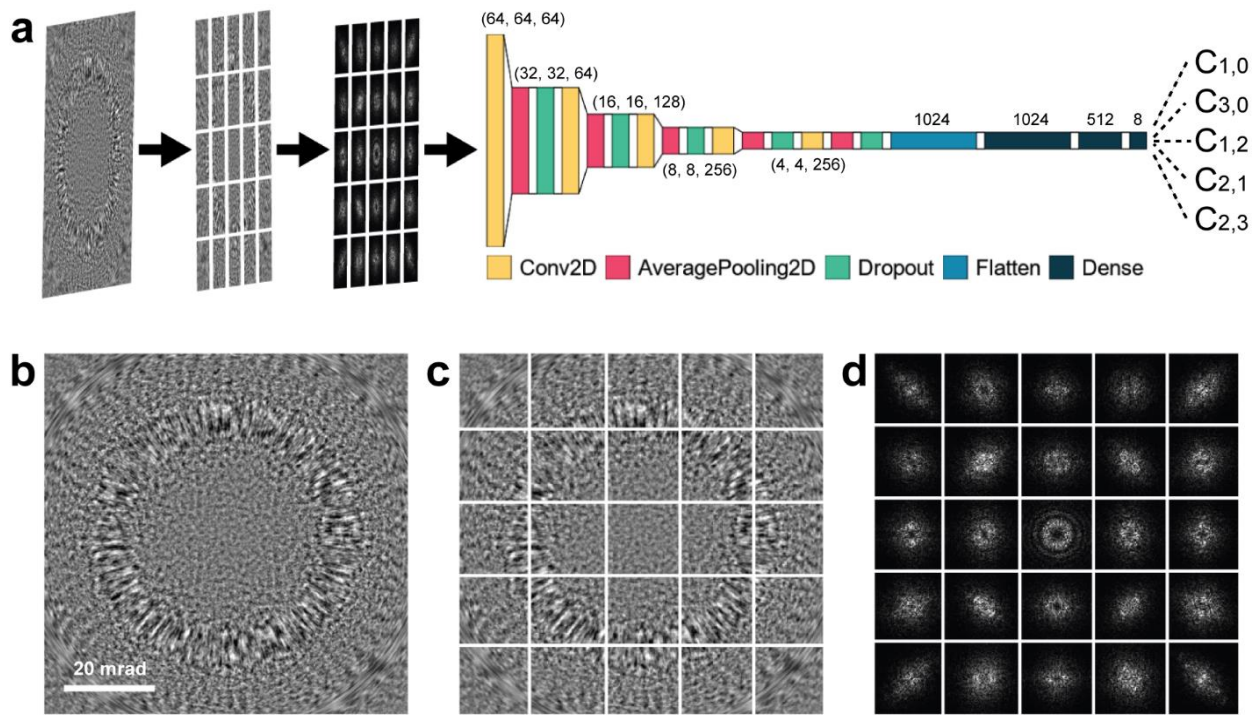

**Figure 3.** (a) Sketch of the artificial neural network (ANN) used to fit synthetic Ronchigrams from a 5 × 5 set of FFTs diffractograms. The ANN consists of 4 sequential blocks of a 2D convolution layer (yellow) followed by a pooling layer (red) and a dropout layer (green). After these blocks, a flatten layer (light blue) was used, followed by 3 dense layers (dark blue) to output the fitted aberration

vales. (b-d) Example of a simulated Ronchigram used as a single input image for the ANN. (b) 2048 × 2048 pixels defocused Ronchigram generated according to Eq. 1. (c) 4 × binned Ronchigram divided into 5 × 5 patterns (or tiles), each of which contains 64 × 64 pixels. (d) Corresponding FFTs diffractograms from the tiles in (b). For clarity, in this example the value of $C_{3,0}$ (spherical aberration) was exaggerated, in order to show its effect on the FFTs in (d) ($C_{3,0}$ = 1 mm).

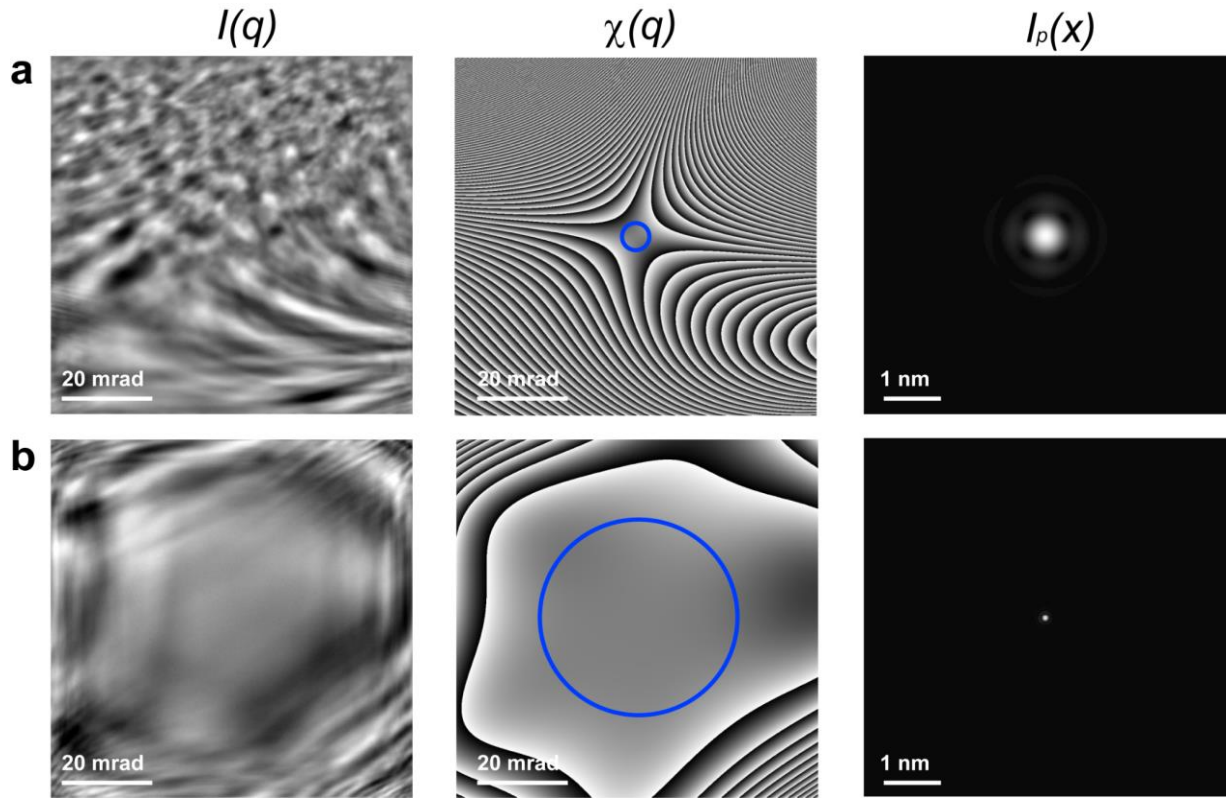

**Figure 4.** Simulated images illustrating the ANN correction capability. (a) Uncorrected probe. (b) Corrected probe after subtraction of the aberration values measured using the ANN. The (Left) Ronchigram intensity $I(\boldsymbol{q})$ and (center) corresponding aberration phase $\chi(\boldsymbol{q})$ calculated at a defocus $C_{1,0}$ = 0 nm. The blue circle indicates the optimal probe aperture $\alpha$ used to estimate the resolution according to the Rayleigh criterion. (Right) Corresponding calculated probe intensity $I_p(\boldsymbol{x})$ in vacuum.

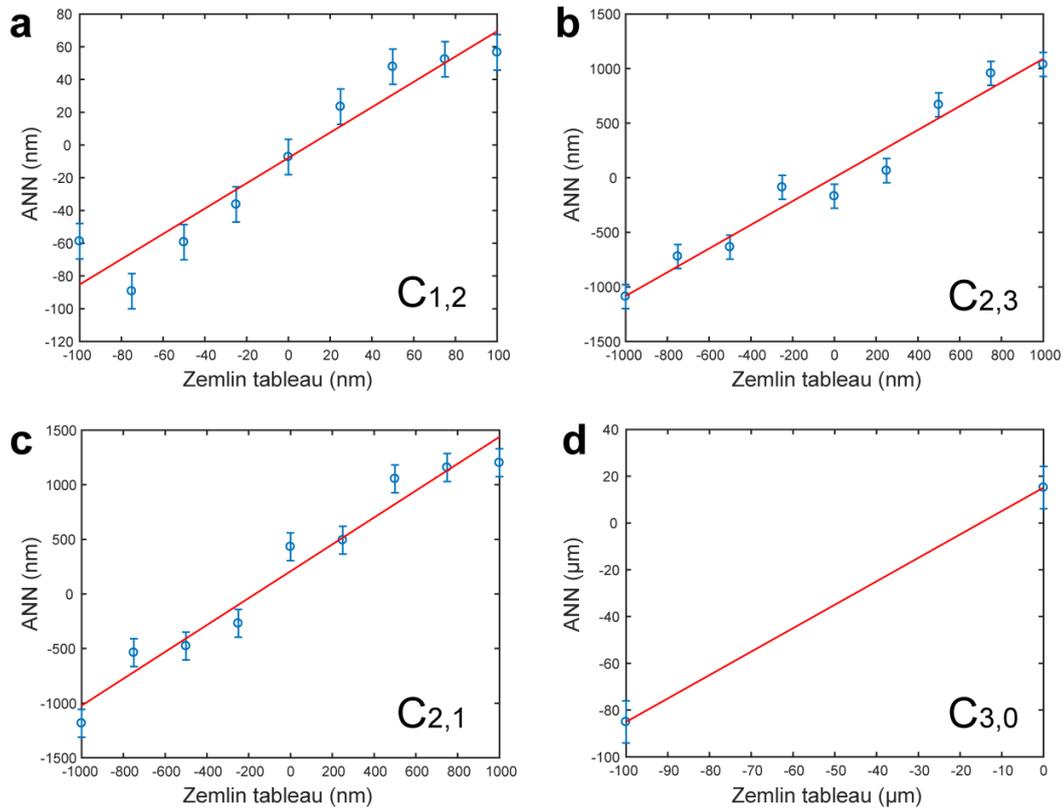

**Figure 5.** (a-d) Comparison between experimental aberration values measured using the ANN and values given by the Zemlin tableau method used by the software controlling the aberration corrector of the STEM probe on the microscope. The circles are experimental data. Error bars are calculated as 95% confidence intervals from the MAE values in Table 1. Red lines are fits to the experimental measurements using a simple 1$^{st}$ order polynomial. The ANN determined the signs of the aberrations in all cases.